# PayPal in Romania

**Teaching Assist. Florentina Anica Pintea,
Teaching Assist. Georgiana Petruţa Fîntîneanu,
Univ. Instr. Bogdan Ioan Selariu
„Tibiscus" University of Timişoara, România**

ABSTRACT. The present paper refers to the usefulness of online payment through PayPal and to the development of this payment manner in Romania. PayPal is an example of a payment intermediary service that facilitates worldwide e-commerce.

## 1. What is PayPal

PayPal is the leader of online paying solutions. It offers individuals as well as legal entities that have an e-mail address, the possibility of making and receiving online pays, easily, fast and above all safely. The service is built on an already existing financial infrastructure of bank accounts and credit/debit cards. To function in complete safety, in real time and on a global level, PayPal utilizes the most advanced system of fraud prevention (proprietary system).

The PayPal.com service functions as a bank account, the only difference being that in a PayPal account everything is done electronically, in front of the computer, via an Internet connection. A user that opens a PayPal account can "fill it up" with an amount, using a bank card compatible with online paying. The money will be withdrawn from the traditional account and deposited in the electronic account.

It is important to know that this operation does not perceive any extra fee, the money transfer being in its self an online payment, banks worldwide having a commission of 0%(zero). Paying on web sites from a PayPal account reduces to a minimum the possibility of fraud in the user's favor /disadvantage.

This site facilitates the Romanian users in making online money transfers to/from users of the same site, as well as in paying for products bought from hundreds of sites which accept PayPal.

A financial consultant says that: "The Romanians have certain reserves in using cards to pay on sites, and PayPal comes to their assistance in the fact that users don't have to disclose their personal data every time".

Via PayPal we can: pay for goods bought online, accept credit cards, send and receive money worldwide, keeping the information related to our card a secret, administrating online expenses.

## 2. Short history about PayPall

Launched in 1998 and bought by Ebay.com in 2002, the Internet paying service PayPal was intended for the use of the American citizens, extending rapidly to the countries of the European Union, as well as Switzerland and Norway. Romania and Bulgaria were held in "stand-by", till September 2007. PayPal extended its services worldwide, over 90 nations having access to this service. Among the new entries into the sites coverage there are: Albania, Bosnia and Herzegovina, Bulgaria, Morocco, San Marino, Saudi Arabia and The Vatican. Among the novelties it is the possibility of using the service in Spanish, French and Chinese.

"We are extremely proud to have 190 countries and versions in the four most spoken languages worldwide", declared, cited by Reuters, Dana Stalder, vice president of PayPal. The accounts opened by the Romanian users have a series of restrictions. For the time being, once the electronic account is charged, the user can transfer money to other PayPal accounts or they can pay for various merchandises bought on the internet. He may not receive money or ask the bank to emit a check in value of a certain amount of money from the account. The officials of the service declared that they would implement these services in time.

## 3. Free PayPal account

Yes, you can create a PayPal account free of charge, to send money to friends and family. You can transfer money from your bank account to your PayPal account free of charge. There are no fees for clients transferring money from PayPal accounts, but there is a limit for the amount of money that can be paid, received or withdrawn from your personal account. There are several types of accounts that vary as to the fees and the maximum sums that can be transferred. In the beginning you can withdraw a maximum sum of 500 Euro per day (depending on the account type). Clients outside of the U.S. will pay a nominal fee in the beginning.[3]

## 4. Why PayPal?

PayPal is the best in online transactions. The service offered by this company is used worldwide, by over 8 million users and dealers. Besides, there are a lot of dealers on the Internet which are not trustworthy. If you disclose details about your credit card you may expose yourself to a great risk. By using PayPal to pay for the merchandise they will not have access to the personal information of your credit card. PayPal keeps that information confidential. They extract money from your card and make payments for the products you buy without revealing information about your card. Using this service you ensure that your personal information stays undisclosed, you pay easily using only your e-mail address.

More and more Romanian sites are implementing systems for online payments, via card, of the products they are trading. Regardless you pay phone bills or fines, this system of payment will "exonerate" you from waiting in cues.

Books, magazines, newspapers, music, clothing, flowers and various electronic products are only a few examples of products which can be bought from Romanian sites by someone which owns a card capable of making online payments. Not only the fact that they are available 24/7, but they are faster and more convenient than traditional methods. Moreover, the customer has a greater intimacy, because there are no clients in front of the PC. Although some inconveniences are involved such as the fact that the customer cannot try the product.

Most online stores trade electronic goods. You can buy small things like computer memory modules and mobile phones, but also plasma TV's with 160 cm diagonals.

E-facturi.ro offers the possibility of paying phone bills, gas, internet and web hosting. After creating an account, you choose the provider you wish to pay and you introduce the data of the bill and the card.

The PayPal virtual bank was invented to avoid unveiling the card's data.

## 5. How to use PayPal

Any Romanian site can receive payments via PayPal. The money obtained from payments or even donations can be withdrawn using a card accepted by the system.

What type of debit cards does PayPal accept for withdrawals?
- ➢ Eminent bank: BCR, ING, Raiffeisen, Piraeus Bank
  Card type: Debit
  Model: Visa Electron, Visa Business
- ➢ Eminent bank: BRD
  Card type: Debit
  Model: VISA Hagi Limited Edition
- ➢ Eminent bank: OTP
  Card type: Credit
  Model: MasterCard
- ➢ Eminent bank: Alpha Bank
  Card type: Debit
  Model: VISA

- The clients navigate on the site, and when they decide to buy an item they press the button for buying via PayPal.
- They pay on the secured pages of PayPal
- They come back to the online store after they paid

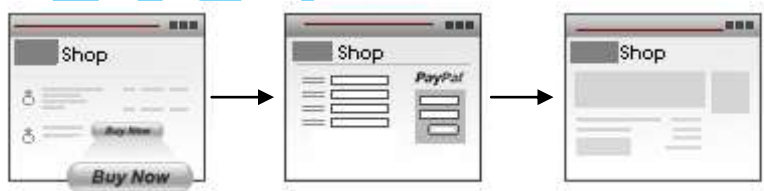

Figure 1

## 6. How can I sign up to PayPal

When a person creates a PayPal account, he has to declare the physical address he lives at. He may not ask the salesman to deliver the goods to another address, because PayPal guarantees only for the address declared at PayPal. If the salesman delivers to another address, PayPal will not offer any warranty if the client claims that the product failed to arrive and

280

demands his money back. This way the client can have the merchandise/product and the money, while the salesman will be the one that supports the loss.[5]

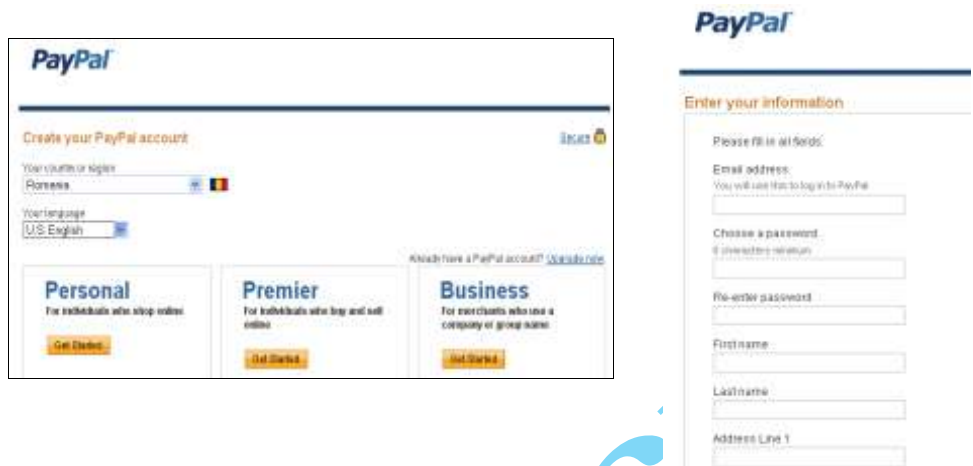

Figure 2

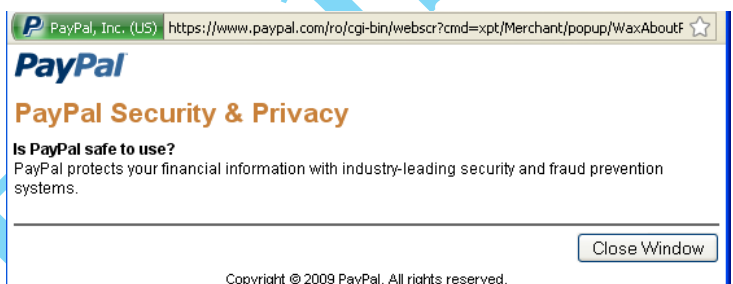

Figure 3

In order to be able to benefit of all the advantages, you will have to follow a simple procedure of confirming a credit card. The verification takes a few days and it is done via the payment of a small sum of 1.5 €, returned on the first acquisition. After this, you have access to a large number of dealers which accept PayPal including eBay.

## 7. Romanian stores that accept PayPal

- USBmania.ro – funny gifts, USB gadgets, office accessories.

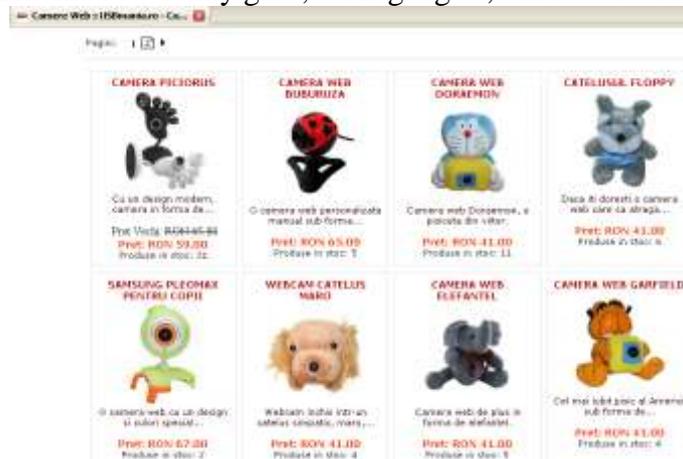

Figure 4

- www.mymovie.ro

Figure 5

- OFIX – online toy, books, music and movie store
- FotoPrint™ - online processing and development for digital photography, frames and photo albums, cups and printed T-shirts
- Voinic.ro – digital cameras, video cameras, Notebooks, DVD players, MP3 players, batteries, memories, lenses

- ➢ PromoTurism – online tourism agency and interactive reservation system, tourism services, seaside bookings, tourist circuits, vacations and plane tickets
- ➢ KERIGMA.ro – Bibles, Christian literature and music, gifts, video/DVD, preaches
- ➢ Omega Shells - Your Ultimate Shell Resource – Home Art Extension
- ➢ Computers, systems, peripherals and PC components, PC Garage Click Phone – internet telephone service!
- ➢ www.florarieinbucuresti.ro

## 8. Internet connection security

The data on a computer's hard-drive is often more valuable that the computer itself, this is why we must give a lot of consideration to keeping it secure. On the other hand even if we don't have extremely valuable data on our hard-drives the risk of infection with a computer virus or another "malware" type program (short for malicious software, harmful programs) can lead to a reduction of system performance or even impossibility of running some software.

PayPal, acquired by eBay in 2002, presently administrates over 133 million accounts worldwide, among witch China, payments in 2005 being evaluated at 27.5 billion $. PayPal and eBay are at the top of the list when it comes to attacks for ID theft.[4]

### 8.1 Computer viruses

In the last years the internet has become the most used medium for spreading computer viruses. Most PC contaminations are done through infected email attachments or infected files downloaded from the internet.

A computer virus is a program that stand out through it's destructive potential of the infected computer. Similar to biological viruses (the flue, etc), computer viruses multiply creating copies of themselves to infect other computers. Contamination with a computer virus is dome by executing an infected file (double clicking on it). In this way the virus is activated and it can begin its destructive process witch can vary extremely in it's harmfulness.[4]

## 9. PayPal developed a hardware security key for it's clients

In the RSA conference held in San Francisco, the online payment company PayPal announced that it will offer it's users a hardware key to ensure a plus of security for the accounts.

The PayPal security key generates a six digit code every 30 seconds and displays the code if the user presses a button. For 5 $, private users will receive a key witch they can register to protect their account. Companies are exempt from paying this tax.

Michael Barrett, PayPal CISO (chief information security officer), declared that the beta variant launched in the U.S.A. is the latest measure adopted by the company to protect it's clients against online fraud.

Clients using PayPal in the USA, Australia, Austria, Canada and Germany will be able to authenticate themselves in their accounts using cell phones.

The service is called PayPal SMS Security Key and uses mobile phones to carry out transactions on 30 sites that use the VIP service offered by the VeriSign certification company. The new service will eliminate the necessity of a password for PayPal, the authentication being made via randomly generated passwords sent through cell phone at the customer's demand.

The company announced that this service will be extended to other European countries in the near future.[4]